\documentstyle[12pt,psfig,wrapfig]{article}
\def\snw{\sin^2\theta_W}

\def\deg{$^\circ$}
\def\etal{{\it et al.}}
\def\gtorder{\mathrel{\raise.3ex\hbox{$>$}\mkern-14mu
             \lower0.6ex\hbox{$\sim$}}}
\def\ltorder{\mathrel{\raise.3ex\hbox{$<$}\mkern-14mu
             \lower0.6ex\hbox{$\sim$}}}

\begin{document}
\begin{center}
{\large \bf Experimental Status of Parity Violating Electron Scattering
\\ }
\vspace{5mm}
E.J.~Beise\\
{\small\it for the SAMPLE collaboration \\}
\vspace{5mm}
{\small\it
University of Maryland, College Park, MD, USA
\\ }
\end{center}

\begin{center}
ABSTRACT \\

\vspace{5mm}
\begin{minipage}{130 mm}
\small

Recently, there has been considerable theoretical interest
in determining strange quark contributions to hadronic matrix 
elements. Such matrix elements can be accessed through the nucleon's
neutral weak form factors as determined in
parity violating electron scattering. A program of experiments is
presently underway or planned at the MIT-Bates, Mainz and Jefferson
Laboratories. The SAMPLE experiment at MIT-Bates 
will measure the strange magnetic form factor $G_M^s$ at low
momentum transfer. Two data taking periods have recently been completed,
representing about 15\% of the desired data on hydrogen. A summary 
of recent progress on
the SAMPLE experiment is presented, along with future plans at Jefferson
Lab.

\end{minipage}
\end{center}

\begin{center}
INTRODUCTION

\end{center}

Elastic electron scattering has been used as a probe of nucleon
structure for many years. The electromagnetic properties of the proton
are very well known, and recently there has been considerable
progress in measuring the electric and magnetic form factors of the
neutron [1,2].  
Additional and complementary information on nucleon structure can
be obtained through the use of neutral weak probes [3]. 
There has been much recent theoretical interest in the possibility
that sizeable strange quark contributions to nucleon matrix elements
may exist. The two most cited pieces of experimental evidence 
are measurements of the $\pi$-nucleon $\Sigma$ term, from which the 
scalar matrix element $\langle N\vert\overline{s}s\vert N\rangle$ 
can be obtained [4], and 
measurements of the nucleon's spin-dependent structure functions in 
deep-inelastic lepton scattering [5], from which the axial 
current matrix element $\overline{s}\gamma_{\mu}\gamma_5 s$ is extracted. 
In each case the $s$-quark contribution to the proton is about 10-15\%,
although both results are sensitive to theoretical interpretation.

With parity violating electron scattering it is possible to 
investigate the vector matrix element 
$\langle N\vert\overline{s}\gamma_\mu s\vert N\rangle$ through
a measurement
of the neutral weak form factor of the proton, {\it i.e.}, the
interaction between a proton and electron through the exchange
of a $Z$ boson [6]. The electromagnetic and weak form factors 
can be constructed as a sum of individual quark distribution 
functions multiplied by coupling constants given by the Standard 
Model for Electroweak Interactions. The lepton currents are
completely determined, and the hadronic currents are the
information to be extracted by experiment. The electromagnetic
coupling gives the well known Sachs form factors $G^{p,n}_{E,M}$.
Neglecting quarks
heavier than the $s$-quark and making the assumption that the
proton and neutron differ only by the interchange of $u$ and $d$
quarks, the neutral weak vector form factors for the proton 
can be expressed in terms of the EM form factors in the following way:
\begin{equation}
G^{Z,p}_{E,M} = \left({1\over 4} - \sin^2\theta_W\right)
  \left[1+R^p_V\right]G^p_{E,M} - {1\over 4}\left[1+R^n_V\right]G^n_{E,M}
   - {1\over 4}\left[1+R^s_V\right]G^s_{E,M}
\end{equation}
The factors $R^i_V$ are weak radiative corrections which must
be applied to account for higher order processes.
In addition the axial vector coupling leads to
\begin{equation}
G^Z_A = -{1\over 2}\left[1+R_A^{T=1}\right]g_A\tau_3 
      + {\sqrt{3}\over 2}R_A^{T=0}G^{(8)}_A 
      + {1\over 4}\left[1+R^s_A\right]G^s_A
\end{equation}
where $\tau_3$=+1(-1) for the proton(neutron).

The term involving the SU(3) isoscalar form factor $G^{(8)}_A$ is generally
ignored since it turns out that it 
is not present at tree level, and an estimate of 
$R^{T=0}_A$~[7] shows that it is suppressed relative to the 
dominant first term.
The isovector axial form factor $g_A(0)$=1.26 is determined from neutron
beta decay.  With the exception of $G^n_E$, the electromagnetic 
form factors are determined with good precision. The axial 
strange form factor $G^s_A$ is the same
quantity extracted from polarized deep inelastic lepton scattering.
The only undetermined
quantities are the strange quark contributions $G^s_{E,M}$. 
At $Q^2$=0, $G^s_E$=0 because the proton has no net strangeness. 
The magnetic form factor $G^s_M(0)$ is not well constrained, and 
the $Q^2$ dependence of all three strange form factors is 
unknown. This has stimulated a program of parity violation
experiments at Bates~[8] Jefferson Lab (formerly CEBAF)~[9] 
and Mainz~[10]. These
experiments represent the first genertion attempt at determining 
the matrix element 
$\overline{s}\gamma_\mu s$ in the proton, and many theoretical 
predictions have been put forth to estimate its size.
Near $Q^2$=0, it can be  characterized by 
two parameters: the ``strange magnetic moment''
$\mu_s = G^s_M(0)$, and the Sachs ``strangeness radius''
$r^2_s = -{1\over 6}{d G_E^s\over d Q^2}$. 
Predictions of various models are listed in table I. 
A notable point  is that although several models have similar predictions
for the magnitude of $\mu_s\sim -0.3$, the 
estimates of $r^2_s$ vary widely depending on the type of mechanism 
assumed.
\begin{center}
\begin{table}
\caption[{\bf\small Table I}]{\small Theoretical predictions for $G^s_M(0)$
and the Sachs radius $r^2_s$, after Musolf {\it et al.}, [7] with some
recent additions.}
\begin{tabular}{lll}
\\
Type of Calculation (reference) & $\mu_s$  & $r^2_s$ (fm$^2$)  \\
\hline \\
Poles~[11] & $-0.31\pm 0.009$ & $0.14\pm 0.07$  \\
Poles~[12]  & $-0.24\pm 0.03$ & $0.21\pm 0.03$ \\
Kaon Loops~[13] & $-0.40\longrightarrow -0.31$ & $-0.03\pm 0.003$ \\
Kaon Loops~[14] & $-0.03$ & $-0.01$  \\
``Loops and Poles''~[15] & $-0.28\pm 0.04$  & $-0.03\pm 0.01$ \\
SU(3) Skyrme~[16] & $-0.33\longrightarrow -0.13$ 
   & $-0.11\longrightarrow -0.19$  \\
SU(3) chiral hyperbag~[17] & $0.42\pm 0.30$ &  \\
SU(3) chiral color dielectric~[18] 
   & $-0.40 \longrightarrow -0.03$ & $-0.003\pm 0.002$ \\
Chiral quark-soliton~[19] & $-0.45$ & $-0.17$ \\
Constituent Quarks~[20] & $-0.13\pm 0.01$  & $-0.002$ \\
``QCD Equalities''~[21] & $-0.73\pm0.30$ &  \\
\hline \\
\end{tabular}
\end{table}
\end{center}

Parity violating electron scattering is evaluated in terms 
of the asymmetry
in the cross section for the scattering of right- and left-helicity 
electrons from an unpolarized target.
For elastic scattering from a free proton, the asymmetry consists of 
three terms which reflect the interference between the NW
and EM interactions:
\begin{equation}
A_p = \left[{G_F Q^2\over \sigma_p\pi\alpha\sqrt{2}}\right]
   \left[ \varepsilon G_E^p G_E^{Z,p}   + \tau G_M^p G_M^{Z,n} 
   - {1\over 2}\left(1-4\snw\right)\varepsilon^{\prime} G_M^p G_A^{Z,p}\right]
    \, , 
\end{equation}
where
$\varepsilon = \left(1+2(1+\tau)\tan^2{\theta\over 2}\right)^{-1}$ 
and 
$\varepsilon^\prime = 
\sqrt{\left(1-\varepsilon^2\right)\tau\left(1+\tau\right)}$.
The elementary unpolarized cross section is proportional
to  $\sigma_p = \varepsilon\left(G^{p,n}_E\right)^2 + 
\tau\left(G^{p,n}_M\right)^2$. 
The kinematic factors result in the first two terms of 
the asymmetry dominating at forward angles and the latter
two terms contributing at backward angles, although the term containing the 
axial vector form factor $G_A^Z$ is suppressed by the factor 
$(1-4\snw)$. 

\begin{center}
THE SAMPLE EXPERIMENT at MIT-BATES

\end{center}

SAMPLE~[8] is a measurement of $A_p$ at backward
angles ($130^\circ < \theta < 170^\circ$) and $E_{lab}$ = 200 MeV, 
resulting in $Q^2 = 0.1$ (GeV/c)$^2$. At these kinematics the parity 
violating asymmetry is sensitive to $G^s_M(0)$=$\mu_s$ and, if 
$\mu_s$=0,  is  $-7.6\times 10^{-6}$ (-7.6~ppm).
The goal of the experiment is to achieve a 
an absolute experimental error of $\delta \mu_s \sim 0.22$, which requires
$\sim$150 Coulombs of polarized beam on a 40~cm liquid hydrogen
target.
Elastically scattered electrons are detected in 
the backward direction by a large solid angle air
{\v C}erenkov detector consisting of ten mirrors which image the
target onto ten 8~inch photomultiplier tubes, as shown in figure 1. 
The average beam current is typically 40~$\mu$A, delivered 
with a 1\% duty cycle at 600~Hz.
The counting rate in each photomultiplier tube is very high, so
individually scattered electrons are not detected but the signal
is integrated over the 15~$\mu$sec beam pulse and normalized
to the charge in each burst. Background is measured 
by closing shutters in front of the phototubes and
with empty target runs. 
It is also necessary to determine the contribution to the light yield which
does not come from Cerenkov light, which is performed in two ways. First,
runs are taken with the mirrors covered leaving the phototube shutters
open. Additionally, data is taken with 
nine out of ten pulses from the accelerator reduced by several 
orders of magnitude so that individually scattered
electrons can be detected in coincidence with photons incident
on the phototube. The tenth pulse is used as a ``tracer bullet'' to directly 
compare the ordinary detector signal to the ``pulse-counting'' signal.

The two most dominant sources of light-producing 
background are (a) scintillation
light generated in the air from particles producing EM 
showers in the target and
(b) Cerenkov light arising from Michel positrons due 
to $\pi^+$ photoproduction
in the target. (Due to the low beam energy, inelastically scattered 
electrons are a negligible contribution.)  The former source is
assumed to have zero asymmetry because they are low-momentum transfer 
EM processes.
The asymmetry of the latter source has been estimated to be very
small ($\sim 1 \times 10^{-7}$) [22]. At present it appears that
elastic scattering comprises at least 50\% of the light yield. A detailed
breakdown of these contributions will result from further simulations
and data analysis which are currently in progress. Additional 
shielding should remove the Michel positrons and thus increase the
relative contribution of elastic events to the light yield in future runs.

%
%
\begin{figure}
\begin{center}
\strut\psfig{figure=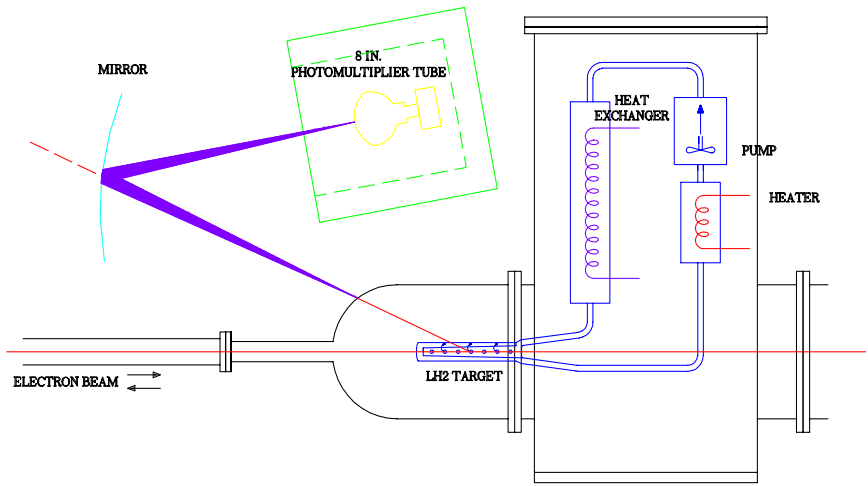,height=2.5 in,width=4.5 in,angle=90}
\caption{\small Layout of the SAMPLE detector.}
\end{center}
\end{figure}
%
%

Circularly polarized laser light from
a Titanium-sapphire laser is incident upon a bulk GaAs crystal,
from which an electron beam of approximately 35\% polarization 
is extracted. The circularly polarized light is generated by 
a linear polarizer followed by a Pockels cell which acts as a 
quarter wave plate when the appropriate voltage is applied.
The helicity of the electron beam is flipped by reversing the
polarity of the voltage on the Pockels cell.
The beam helicity is chosen randomly
on a pulse-by-pulse basis, except that pulse pairs 1/60 sec
apart have opposite helicity. ``Pulse-pair'' asymmetries are formed
every 1/30 sec, greatly reducing sensitivity to 60~Hz electronic
noise and to drifts in beam properties such as current, 
energy, position and angle. In addition,
a $\lambda$/2 plate can be manually rotated upstream of the 
Pockels cell to manually reverse the polarization of the beam
independent of all electronic signals.

The electron beam deposits approximately 550 watts of power into 
the liquid hydrogen target. Density fluctuations are minimized by subcooling
the liquid below its boiling point by a few degrees and by
rapidly circulating the fluid in a closed loop such that a 
packet of hydrogen is in the path of the beam for only a short
time. No density reduction was seen in the normalized yield at 
the level of a few percent as the beam current was varied
between 4 to 40~$\mu$A. Fluctuations in the normalized yield due to
variations in beam properties limit the accuracy with which 
this observable can be used to determine density changes. A
more sensitive determination is to monitor the width of
the pulse-pair asymmetry, and in this observable no
change in density was seen at the level of much better than 1\%~[23].

If helicity correlations are present in the beam properties they will
cause false asymmetry contributions to the data.
The raw measured asymmetry $A_{raw}$ must be corrected for these effects.
The corrected asymmetry $A_c$ can be expressed as
\begin{equation}
A_c = A_{raw} - {1\over S} \Sigma_i {\partial S\over\partial\alpha_i}
  \delta\alpha_i^{LR} \, ,
\end{equation}
where $S$ is the normalized detector 
yield, $\alpha_i$ is one of
five beam parameters (position and angle in $x$ and $y$, and energy),
and $\delta\alpha_i^{LR} = \alpha_i^R - \alpha_i^L$ is the helicity
correlated difference in the beam parameter. 
To first order, the experiment is designed to be as insensitive
as possible to fluctuations in beam properties. This includes
feedback loops to stabilize the beam energy and position on target,
as well as a feedback loop between the beam charge asymmetry in the 
accelerator and the Pockels cell voltage. With this, 
the helicity correlations due to the beam intensity are
reduced from 100-200 ppm 
to $\sim$1ppm or less.
The corrections to the asymmetry are then made with 
measured helicity correlations in the beam and the measured detector sensitivity 
to beam properties as determined from the natural long term drifts in the beam.
Other properties such as helicity correlated differences
in beam size are studied with dedicated runs in between data taking and
appear to be negligible. 

Two running periods have now been completed for SAMPLE, one in fall 1995 
and the other in spring 1996, totalling about 21 Coulombs of good data.
(requiring any individual correction to the data be less than 1 ppm). 
Figure 2 shows
the asymmetry in (a) beam current, (b) horizontal beam position for all good data. 
The open symbols (``NORMAL'') are runs with the normal orientation of the 
$\lambda$/2 plate and the closed points (``REVERSE'') are with the 
helicity manually reversed. 
In Panel (c) is the raw detector asymmetry, and in (d) the asymmetry after
corrections due to helicity correlations in the beam are applied.

%
\begin{figure}
\begin{center}
\strut\psfig{figure=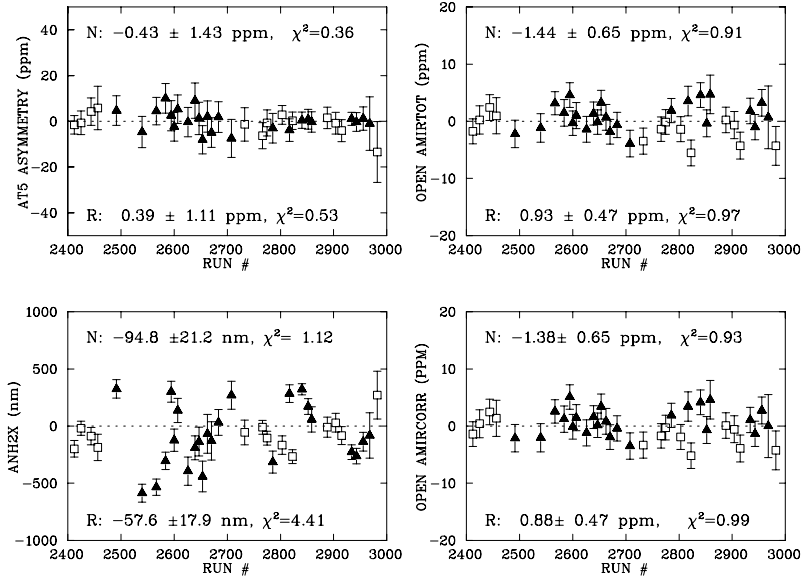,height=3.0 in,angle=90}
\caption{\small
Measured asymmmetries in (a) beam intensity and (b) horizontal
beam position, and the  detector asymmetry (c) before and (d) after corrections
due to helicity correlated beam properties are applied.}
\end{center}
\end{figure}

The SAMPLE
beam line is at an angle with respect to the accelerator, which 
will cause the electron spin to precess by 16.7{\deg} from longitudinal.
The beam spin can be made longitudinal at the SAMPLE target by preparing
the spin off-axis at the injector with a Wien filter. Nonetheless, any
remaining small transverse component in the beam will 
result in a parity conserving Mott
asymmetry if the detector is not perfectly azimuthally symmetric. The Mott
asymmetry was measured by  preparing the electron beam in its two possible transverse
states ($\phi$=0{\deg} and $\phi$=90{\deg}). The Mott contribution 
to the full detector physics asymmetry was determined to be 
consistent with 0 at the level of 0.5~ppm. 

Analysis of the 1996 data set is presently underway. 
Combining all of the good data from 1995 and 1996, each shown in 
figure~3,  results in a physics
asymmetry that is approximately of the same magnitude as the expected 
PV asymmetry of about -6 ppm. At present the
statistical error on the data results in an error on $\mu_s$ of $\pm 0.4$.

At the SAMPLE kinematics the contribution from the axial vector
form factor term $G_A^{Z}$ is about 20\%. The weak radiative correction
to this term $R^{T=1}_A$ has been estimated to be $-0.34\pm 0.34$ [24]. 
This leads to a theoretical limit on the
ultimate uncertainty in the experimental result corresponding
to $\delta G^s_M \sim\pm 0.18$. A direct measure
of this radiative correction would therefore be useful. This can 
be achieved by measuring the asymmetry in
quasielastic scattering from deuterium at the same kinematics, and
taking the ratio $A_p/A_d$ would as a result reduce
the theoretical error to $\delta G^s_M\sim\pm 0.02$. Running
with deuterium is planned in the future. 

\begin{figure}
\begin{center}
\strut\psfig{figure=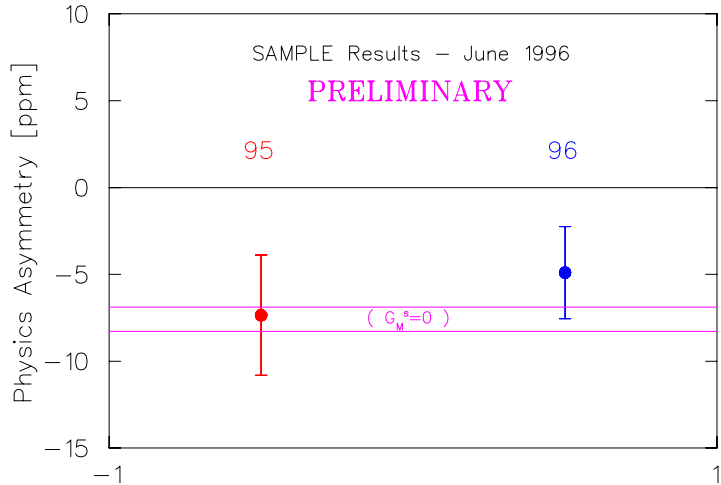,height=2.5 in,angle=90}
\caption{\small
Preliminary results for the parity violating asymmetry from the 
1995 and 1996 SAMPLE running periods. The error bars include an estimate
of the systematic uncertainty due to background processes.}
\end{center}
\end{figure}

\begin{center}
JEFFERSON LAB EXPERIMENTAL PROGRAM

\end{center}

At higher beam energies and forward angles one has the possibility of
obtaining information on the $Q^2$ dependence of the strange form
factors, although an explicit extraction of $G_E^s(Q^2)$ from
$e$-$p$ scattering also requires good knowledge of neutron electromagnetic
properties. 
Measurements of PV $e$-$p$ scattering are planned for Jefferson
Lab and Mainz. (For a description of the latter, see the contribution
by F.~Maas to these proceedings.)

Experiment E91-010 at Jefferson Lab will measure $A_{PV}$ in the proton  
at $Q^2=0.5$ (GeV/c)$^2$ using the two high resolution 
spectrometers in Hall A. 
The standard detector package of the Hall A spectrometers will be replaced
with a shower counter, with integrating electronics in order to allow
for the very high expected rates into the spectrometer. 
This measurement alone would not allow 
separation of $G_E^s$ and $G_M^s$, but it would determine
if strange quarks play a significant role in the structure of the proton,
modulo the uncertainty in $G_E^n$. If large effects are seen, 
a more extended program of measurements at lower $Q^2$
on proton and $^4$He targets is anticipated.

The ``G0'' experiment, E91-017, will consist of a superconducting
toroidal spectrometer and an array of scintillators along the focal
plane to determine the PV asymmetry at both forward and backward
electron angles.  At forward electron kinematics the detector will 
be sensitive to 
recoil protons corresponding to 
momentum transfers of $0.1<Q^2<1.0$ GeV$^2$.
A second set of measurements in which the orientation of the
spectrometer is reversed will allow the detection of electrons at 108{\deg}.
Additional measurements on deuterium will
constrain the uncertainty associated with $G_A^Z$.
$G_E^s$ and $G_M^s$ can then be separated over a range of momentum transfer. 

Another approach to determining strange quark effects in a
hadronic system is to use parity-violating elastic scattering from
a $J=0$, $T=0$ nucleus. Elastic electron scattering from a 
spinless nucleus occurs only through charge scattering.  
For $T=0$ nuclei, only isoscalar terms remain and  the asymmetry reduces to 
\begin{equation}
A = {G_F Q^2 \over \pi\alpha\sqrt{2}} \left[ \snw + {1\over 4} 
    {F_s \over F_c}\right] \, .
\end{equation}
$F_c$ is the electromagnetic charge response function of the nucleus
and $F_s$ is the equivalent strange quark response function.

Multinucleon effects which could complicate the extraction of $G_E^s$ from
a nuclear target have been calculated [25] and appear to be small.
Musolf and Donnelly have also argued [26] 
that two measurements on a ($0^+0$) nucleus, at low and moderate $Q^2$ 
might ultimately constrain $G_E^s$ to a higher level of precision than
that achievable with a measurement on a proton target, since uncertainties
associated with $G_M^s$, $G_E^n$ and $G_A^Z$ are no longer present.

Experiment 91-004 will measure $^4$He$(e,e^{\prime})$ at
$Q^2=0.6$ GeV$^2$, again using the Hall A spectrometers. At the highest TJNAF 
design luminosity (${\cal L}\sim 3.2\times 10^{38}$), the counting rate into
each spectrometer is sufficiently low to track individual particles
through the spectrometer. The asymmetry with no strange quarks 
is relatively large, ($\sim 50$ ppm), but because of the low counting rates the 
experiment is expected to be statistics limited.  However, even a 
modest measurement of the asymmetry will be able to determine if strange 
quark contributions to hadronic properties are large. 

\begin{center}
CONCLUSIONS

\end{center}

Theoretical interest in understanding the role of strange quarks in 
the nucleon has stimulated a new generation of experiments
in parity-violating electron scattering and also in
neutrino scattering~[27]. Currently very 
little information is known. Considerable progress on the SAMPLE
experiment has been made and a long data-taking run is anticipated
in 1997. Future running with a deuterium target is also planned.
The first experiments using polarized beam at Jefferson Laboratory are
planned for 1997, establishing the groundwork for a broad program
of PV experiments in the future.

Work on the SAMPLE experiment is supported by the National
Science Foundation and by the Deptarment of Energy.

%
%
%
%
\vspace{0.2cm}
\vfill
{\small\begin{description}
\vspace{0.2cm}
\parskip=0pt
\item{[1]}  S.~Platchkov {\it et al.}, Nucl.~Phys.~{\bf A510}, 740 (1990).
   M.~Meyerhoff {\it et al.}, Phys.~Lett.~{\bf B327}, 201 (1994).
   See also  J.~Becker, these proceedings.
\item{[2]} A.~Anklin {\it et al.}, Phys.~Lett.~{\bf B336},
   313 (1994). H.~Gao {\it et al.}, Phys.~Rev.~{\bf C50}, R549 (1994).
   EE.W.~Bruins {\it et al.}, Phys.~Rev.~Lett.~{\bf 75}, 21 (1995).
\item{[3]} D.~Kaplan and A.~Manohar, Nucl.~Phys.~{\bf B310}, 527 (1988).
\item{[4]} J.~Gasser, H.~Leutwyler, and M.E.~Sainio, 
   Phys.~Lett.~{\bf B253}, 163 (1991).
\item{[5]} G.~Mallot, these proceedings, and references therein.
\item{[6]} R.D.~McKeown, Phys.~Lett.~{\bf B219}, 140 (1989); D.H.~Beck,
Phys.~Rev.~{\bf D39}, 3248 (1989).
\item{[7]}  M.~Musolf, T.W.~Donnelly, J.~Dubach, S.J.~Pollock, 
   S.~Kowalski, and E.J.Beise, Phys.~Rep.~{\bf 239}, 1 (1994).
\item{[8]}  Bates experiments 89-06 (R.~McKeown and D.~Beck, contacts), 
and 94-11 (M.~Pitt and E.~Beise, contacts).
\item{[9]} TJNAF experiment E91-017 (D.~Beck, contact), E91-010, (P.~Souder, contact),
and E91-004, (E.~Beise, contact).
\item{[10]}  Mainz proposal \#~A4/1-93 (D.~von Harrach, contact). See also
F.~Maas, these proceedings.
\item{[11]}  R.L.~Jaffe, Phys.~Lett.~{\bf B229}, 275 (1989).
\item{[12]}  H.W.~Hammer, U-G.~Mei{\ss}ner, and D.~Drecshel, 
 preprint TK-95-24, 1995, also e-print \# hep-ph/9509393.
\item{[13]} M.J.~Musolf and M.~Burkardt, 
  Zeit.~Phys.~{\bf C61}, 433 (1994).
\item{[14]}   W.~Koepf, E.M.~Henley and S.J.~Pollock, 
  Phys.~Lett.~{\bf B288}, 11 (1992).
\item{[15]} T.~Cohen, H.~Forkel, M.~Nielsen PL{\bf B316},1 (1993). 
\item{[16]} N.W.~Park, J.~Schechter and H.~Weigel, 
  Phys.~Rev.~{\bf D43}, 869 (1991).
\item{[17]} S.~Hong and B.~Park, Nucl.~Phys.~{\bf A561}, 525 (1993).
\item{[18]}  S.C.~Phatak and S.~Sahu, Phys.~Lett.~{\bf B321}, 11 (1994).
\item{[19]} H.C.~Kim {\it et al.}, preprint RUB-TPII-11-95, also
  e-print \# hep-ph/9506344. 
\item{[20]}  H.~Ito, Phys.~Rev.~{\bf C52}, R1750 (1995).
\item{[21]} D.~Leinweber,  Phys.~Rev.~{\bf D53}, 5115(1996).
\item{[22]} S.P.~Li, E.M.~Henley, and W.Y.P.~Hwang, Ann.~Phys.~{\bf 143},
372 (1982).
\item{[23]} E.J.~Beise {\etal}, Nucl.~Inst. and~Meth.~{\bf A378}, 383 (1996).
\item{[24]}  M.~Musolf and B.~Holstein, {\it Phys.~Lett.}~{\bf B242},
461 (1990), and M.~Musolf, private communication.
\item{[25]} M.J.~Musolf and T.W.~Donnelly, Phys.~Lett.~{\bf B318}, 263 (1993), and
 M.J.~Musolf, R.~Schiavilla, and T.W.~Donnelly, Phys.~Rev.~{\bf C50}, 2173 (1994).
\item{[26]} M.J.~Musolf and T.W.~Donnelly, Nucl.~Phys.~{\bf A546}, 509 (1992).
\item{[27]}  Los Alamos LSND experiment, W.C.~Louis, contact.
\end{description}}

\end{document}